\documentclass[letterpaper]{article}

\usepackage[T1]{fontenc}
\usepackage{lmodern}
\usepackage{geometry}
\usepackage{setspace}
\usepackage{balance}
\usepackage{mathptmx}
\usepackage{lastpage}
\usepackage[format=plain,justification=justified,singlelinecheck=false,font={stretch=1.125,small,sf},labelfont=bf,labelsep=space]{caption}
\usepackage{fancyhdr}
\usepackage{fnpos}
\usepackage{csquotes}
\usepackage[english]{babel}
\addto{\captionsenglish}{%
}
\usepackage{array}
\usepackage{droidsans}
\usepackage{charter}
\usepackage[T1]{fontenc}
\usepackage{setspace}
\usepackage[compact]{titlesec}
\usepackage{hyperref}
\usepackage{xspace}
\usepackage[style = chem-acs,
articletitle=true]{biblatex}
\usepackage{graphicx}
\usepackage{float}
\newfloat{scheme}{htbp}{los}
\floatname{scheme}{Scheme}
\floatname{chart}{Chart}
\newfloat{graph}{htbp}{loh}

\usepackage{chemformula} 
\usepackage[version = 4]{mhchem} 

\usepackage{authblk}
\author[1,2,5,\#]{Hojin Kim}
\author[1,3,\#]{Samantha M. Livermore}
\author[4]{Yongjin Shin}
\author[1,3]{Heinrich M. Jaeger}
\affil[1]{James Franck Institute, The University of Chicago, Chicago, Illinois 60637, USA}
\affil[2]{Pritzker School of Molecular Engineering, The University of Chicago, Chicago, Illinois 60637, USA}
\affil[3]{Department of Physics, The University of Chicago, Chicago, Illinois 60637, USA}
\affil[4]{Department of Semiconductor Convergence Engineering, Dankook University, Yongin-si 16890, South Korea}
\affil[5]{Current address: Research Division, ExxonMobil Technology and Engineering Company, Annandale, New Jersey 08801, USA}
\title{The Role of Hydrogen Bridging Bonds in the Shear-Thickening and Jamming of Dense Suspensions}
\date{*Email: hojinkim718@gmail.com}

\begin{document}

\maketitle

\begin{abstract}
\noindent Strong shear thickening and jamming in dense suspensions are driven by friction as particles are sheared into contact.
Control over these frictional interactions can be achieved via particle shape and roughness, and also via the particles’ surface chemistry and interactions with the surrounding solvent. 
We report on experiments with cornstarch suspensions where friction is enhanced by molecular bridging when hydrogen atoms at the ends of solvent molecules bond with hydroxyl groups on the surfaces of adjacent particles. 
We systematically vary the hydrogen bonding propensity by increasing the size of the backbone of the solvent molecule, from water to diols with up to 4 carbon atoms.
For a fixed particle weight fraction, we find a sudden transition from strong shear thickening (in water and ethylene glycol) to shear thinning (in propanediol and butanediol). 
Combining data from rheology, density functional theory simulations, and fixed-rate pull tests, our results show how changes in the solvent’s molecular structure affect both particle-solvent and solvent-solvent interactions, and how this can be used to tailor the shear thickening and jamming behavior of suspensions.
\end{abstract}

\section{Introduction}
In dense suspensions, an increase in viscosity $\eta$ under applied shear is known as shear thickening \cite{Barnes_1989,Morris_2020,Mari_Seto_Morris_Denn_2015}.
This thickening can be continuous (CST) or discontinuous (DST), depending on the details of particle interactions and the particle volume fraction $\phi$. 
DST, in particular, has been associated with networks of frictional particle contacts, which dynamically evolve under shear and thereby resist flow \cite{brown_shear_2014,Wyart_Cates_2014,Seto_Mari_Morris_Denn_2013, singh2026jamming}.
The formation of these contact networks is driven by a stress-activated transition from lubricated to frictional particle-particle interactions \cite{Morris_2018}.
At sufficiently large $\phi$ and shear stress $\tau$, suspensions can furthermore enter a shear-jammed (SJ) state, in which the contact network imparts solid-like rigidity to the suspension \cite{Seto_Mari_Morris_Denn_2013,Mari_Seto_Morris_Denn_2015,Morris_2018,Van_Der_Naald_Singh_Eid_Tang_DePablo_Jaeger_2024}.
While the frictional particle-particle interactions can be controlled by particle shape and roughness\cite{Hsiao_Jamali_Glynos_Green_Larson_Solomon_2017,Schroyen_Hsu_Isa_VanPuyvelde_Vermant_2019,hsu2018roughness,guy2015towards,krishnamurthy2005shear}, the fact that adjacent particles are sheared into contact implies that their surface chemistry can play a significant role as well.

The surface chemistry of particles in suspension is an important factor in establishing solvation conditions.
A strong solvation layer is known to sterically prevent particle aggregation in the quiescent, zero-shear limit.
Particularly strong solvation arises when solvent molecules can bind to functional groups at the particle surfaces \cite{Raghavan_Walls_Khan_2000}.  
Such solvation layers act as a repulsive barrier and resist the formation of frictional contacts as the applied shear is increased.
However, once the applied shear has become strong enough to nearly close the gap between adjacent particles, suitable solvent molecules can form bonds with both particle surfaces.
Such molecular bridges then give rise to a large effective friction.
The somewhat paradoxical result is that strong solvation can be responsible both for very effective particle dispersion as well as pronounced shear thickening and jamming.

A recent example of this behavior is seen in suspensions of particles with thiol-functionalized brush layers in a solvent of telechelic macromolecules that bind to the brushes to form chemical bridges \cite{ Jackson2022,kim2023dynamic}. 
In these systems, dynamic chemical bonding allows for control of the decay rate of the bridging bonds, making it possible to control the strength as well as the time dependence of the effective friction.
A more general chemistry occurs in the solvation of un-functionalized cornstarch and silica suspensions where hydroxyl groups at the particle surface form hydrogen bonds (H-bonds) with nearby solvent molecules.
For sufficiently large applied shear, this leads to enhanced effective friction via interparticle H-bonding and thereby moves the onset of DST and SJ to lower $\phi$.
\textcite{Raghavan_Walls_Khan_2000} showed in systematic experiments how strong H-bonding between particles and solvent results in the formation of solvation layers that prevent particle aggregation at low stress, while \textcite{james_interparticle_2018} demonstrated how interparticle H-bonds that become activated by strong shear can increase interparticle friction to bring about SJ behavior when otherwise only DST would be observed. 
Similarly, \textcite{Bourrianne_Niggel_Polly_Divoux_McKinley_2022} found that DST in silica suspensions cannot take place unless a critical threshold of interparticle H-bonds is met, i.e., unless the frictional coupling is sufficiently enhanced.

While these prior studies clearly indicate that particle-solvent interactions affect shear thickening and jamming in dense suspensions, there has been no systematic investigation of how these behaviors evolve when the size of the solvent molecule forming the bridging bonds is varied.
Here we report on experiments that tests this size dependence systematically with suspensions of cornstarch particles. 

As our baseline, we start with water as the solvent. 
Such aqueous cornstarch suspensions are the prototypical system where H-bonding generates strong solvation in the quiescent state as well as pronounced DST and SJ at large applied stress.
We then move to longer, diol solvents. 
By definition, these linear molecules have hydroxyl (OH-) groups at both ends, which gives them comparable H-bonding capability to water, while their length can be controlled by adding carbon atoms to the backbone.
Specifically, we use ethylene glycol (two carbons), 1,3-propanediol (three carbons), and 1,4-butanediol (four carbons). 
This series of solvents sidesteps a complication in prior work \cite{van_der_naald_role_2021} where the use of polyethylene glycol not only introduced a longer solvent molecule but also additional H-bonding sites.
The series also spans a range of dielectric constants, meaning that we also vary solvent polarizability.
Given the comparable density $\rho$ of all four solvents (see Table~\ref{tab:solvents}), increasing the length of the solvent molecule reduces the molecular number density $\rho_\mathrm{n}\sim \rho\cdot V/MW$ of H-bonding sites at a fixed volume fraction, where $MW$ is the molecular weight and $V$ is the total volume of solvent.

To relate the macroscopic flow behavior to the molecular-scale interactions of the solvent molecules, we measure the shear rheology of each suspension at several particle weight fractions. 
We complement this with density functional theory simulations that provide insight into the role of molecular conformations of the solvent.
While rheological measurements can investigate the response to applied shear stresses up to around 100~Pa in these suspensions, reaching deep into the shear jammed regime requires significantly larger applied stresses.
To probe both the appearance of shear jamming and, at tens of kPa, the eventual failure of the shear jammed solid, we perform tensile pull tests \cite{james_interparticle_2018,Majumdar_Peters_Han_Jaeger_2017}.
In our tests, an initially submerged rod is pulled vertically upward and the deformation of the suspension under tensile strain is observed with high-speed video.
Compared to extensional rheology such as CABeR \cite{McKinley_Sridhar_2002} for the liquid-like material regime, our tests apply a fixed extensional strain rate and we track the onset of fracture rather than measuring the necking profile as a function of time.
Obervations of the failure mode are then used to infer the effect of solvents on the strength of the shear jammed solid. 
Taken together, this study reveals how small changes in the interaction between solvent and particle surfaces as well as between solvent molecules can profoundly affect the stress response of dense suspensions.

\begin{table}
    \centering
    \begin{tabular}{c|c|c|c}
        Solvent & $\rho$ (g/cm$^3$) \textsuperscript{1} & $\eta$ (mPa$\cdot$s) & $\varepsilon_\mathrm{r}$\textsuperscript{2}\\
        \hline
        Water (W) & 0.89 & 1.00 & 78 \\
        Ethylene glycol (EG) & 1.113 & 19.8\cite{yue_ethylene_2012} & 42 \\
        1,3-Propanediol (PD) & 1.053& 52.0\textsuperscript{1} & 35 \\
        1,4-Butanediol (BD) & 1.017& 84.9\textsuperscript{1} & 31
    \end{tabular}
    \caption{Characteristics of solvents used in this investigation; density $\rho$ at 25$^\circ$C, viscosity $\eta$ at 20$^\circ$C, dielectric constant $\varepsilon_\mathrm{r}$ at 20$^\circ$C.
    \textsuperscript{1}Values provided by suppliers. \textsuperscript{2}Data from Refs.~\cite{maryott1951table,ikyumbur2019dielectric}}
    \label{tab:solvents}
\end{table}

\section{Methods}

\subsection{Suspension preparation}
All suspensions in this study were made with cornstarch particles (Sigma-Aldrich, S4180, $\rho=1.59$~g/cm$^3$) suspended in a single solvent.
The solvents used in this study are water, ethylene glycol (99\%, Alfa Aesar, A11591), 1,3-propanediol (98\%, Sigma-Aldrich, P50404), and 1,4-butanediol (99\%, Sigma-Aldrich, 493732).
\textbf{\textit{Caution!}} 1,4-butanediol (CAS 110-63-4) is considered hazardous according to OSHA regulations due to danger of acute oral toxicity (category 4) and single exposure central nervous system toxicity (category 3).
Skin exposure and inhalation of gases was avoided by using gloves and working in a fume hood.

Using this series of solvents enabled us to probe the role of hydrogen bonding without the formation of complex molecular interactions between the solvent and the particle surface.
We also limited ourselves to pure solvents rather than mixtures of solvents to avoid added complexity from solvent-solvent interactions between species.
The dielectric constant decreases as the number of carbon atoms in the solvent increases, as listed in Table~\ref{tab:solvents}.
Suspensions were prepared by mixing dry cornstarch particles and solvent. 
Particle concentration is reported as a solid weight fraction $\phi_\mathrm{w}$. 

All rheology experiments were performed within 1~hour of sample preparation in order to avoid aging effects that result from the swelling of cornstarch particles in the solvent \cite{chen2019discontinuous}.

\subsection{Shear rheology}
All rheology was measured on an Anton Paar MCR 301 rheometer using a 25~mm parallel plate geometry. 
Viscosity was measured using stress-controlled conditions, where the applied stress was ramped logarithmically from 0.1 to 100~Pa.
Suspensions at packing fractions near the jamming point are particularly sensitive to external stress; small amounts of stress exerted during the sample loading can, therefore, lead to jamming. 
To ensure consistent rheology, a pre-shear step was applied (0.1~Pa for 60~s), and reproducible rheology was verified over several ($\geq3$) measurements of different samples. 
At each stress, the suspension was sheared for 30~s to equilibrate before the steady-state viscosity was measured. 
All rheology was measured within 1~hour of combining particles and solvent, at ambient room temperature.
Error bars in the reported rheology curves are the standard deviation calculated based on at least three separate measurements.

\subsection{Density functional theory simulations}
We performed density functional theory (DFT) calculations with the generalized gradient corrected functional proposed by Perdew‒Burke‒Ernzerhof (PBE) \cite{PBE} and hybrid functional Becke three parameter Lee-Yang-Parr (B3LYP) \cite{B3LYP}, using the plane wave code Quantum ESPRESSO (QE) \cite{Giannozzi_2009,Giannozzi_2017,Giannozzi_2020}.
A kinetic energy cutoff of 60 Ry for the wavefunctions and Fock exchange, and 240 Ry for the charge density ware used.
We used Optimized Norm-Conserving Vanderbilt (ONCV)\cite{ONCV} pseudopotentials, and the calculations were carried out in 20 \AA$\times$20 \AA$\times$20 \AA\xspace unit cells with vacuum with the Brillouin zone sampled by a single $k$-point ($\Gamma$).
The atomic coordinates were optimized with a threshold of 1$\times$10$^{-4}$ Ry/Bohr ($\sim$3 meV$\cdot$\AA$^{-1}$) on the atomic forces. Visualization of the atomic structures was done using the VESTA package \cite{VESTA}.

We computed the phonons of ethylene glycol, 1,3-propanediol, and 1,4-butanediol molecules using density functional perturbation theory \cite{DFPT1,DFPT2}. From these calculations, we obtained the frequencies of the vibrational modes, which we then used to compute the Helmholtz free energies, $F$, as defined by the following equation:
\begin{equation}
\label{eq:Fvib}
F = E -k_BT~\mathrm{ln} Z_n = E -\frac{1}{2}\sum_{i}\hbar\omega_i +k_BT\sum_{i}\mathrm{ln}(\mathrm{exp}^{\hbar\omega_i/k_BT}-1),
\end{equation}
where $E$ is the total energy and $\omega_i$ is the $i$-th vibrational frequency obtained from phonon calculations.
Since the phonon code in QE is implemented for PBE functionals only, we applied the same vibrational entropy term to both PBE and B3LYP total energies. This treatment is justified because PBE-level phonon properties are in agreement with experiments in systems with varying electronic correlations \cite{Shin/Rondinelli2022a,Shin2016,Yan/Shin/Galli/Nocera2023}.

\begin{figure}[ht]
    \centering
    \includegraphics[width=0.6\linewidth]{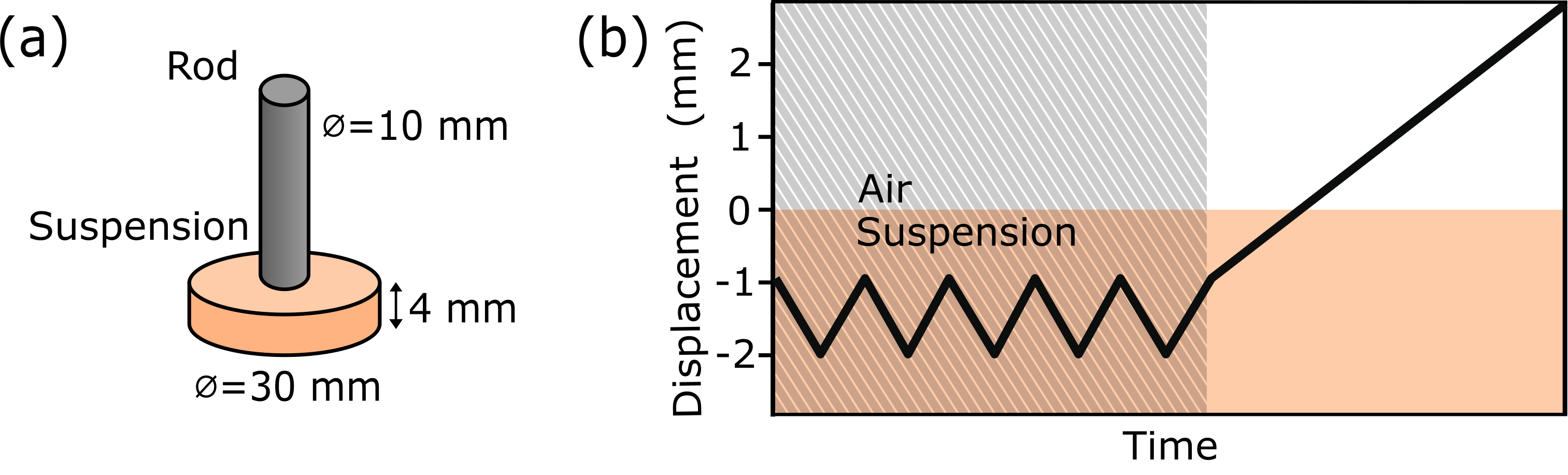}
    \caption{
    Schematic of the pull test setup (a) and the testing protocol (b), which includes a conditioning phase (shown shaded).
    }
    \label{fig:pull_test_schematic}
\end{figure}

\subsection{Pull tests}
For the pull tests a cylindrical, 10~mm diameter aluminum rod was submerged into a small volume of suspension and retracted vertically upwards at speed $v$ using a Zwick-Roell Z1.0 universal tester with 1~kN load cell.
A schematic of the experimental setup is shown in Fig.~\ref{fig:pull_test_schematic}. 
The suspension was confined to a cylindrical well of depth 4~mm and diameter 30~mm.
To improve reproducibility, the aluminum pull-test rod was detergent-cleaned prior to each experiment (Dawn, Proctor \& Gamble) to standardize wettability.
The rod position was initialized at 1~mm below the surface of the suspension at the start of the test.
During an initial conditioning phase, the rod was moved 1~mm further down into the suspension and back up again five times at a fixed speed 1~mm/s to remove any shear-induced jamming that may have occurred during sample loading.
Once the conditioning was complete, the rod returned to its starting position ($z=-1$~mm) before pulling up and out of the fluid at speeds $v$ up to 50 mm/s.
At $v=50$ mm/s, the initial acceleration of our instrument was non-negligible (see Supporting Information and Figure~S1 ); as such, regions of non-zero acceleration are marked clearly in the data presented.
Pull tests were imaged with a Phantom VEO high-speed camera equipped with an Ex Sigma DG Macro lens (focal length 105~mm).

\section{Results and Discussion}
\subsection{Low-stress shear rheology}

\begin{figure*}[ht]
    \centering
    \includegraphics[width=\linewidth]{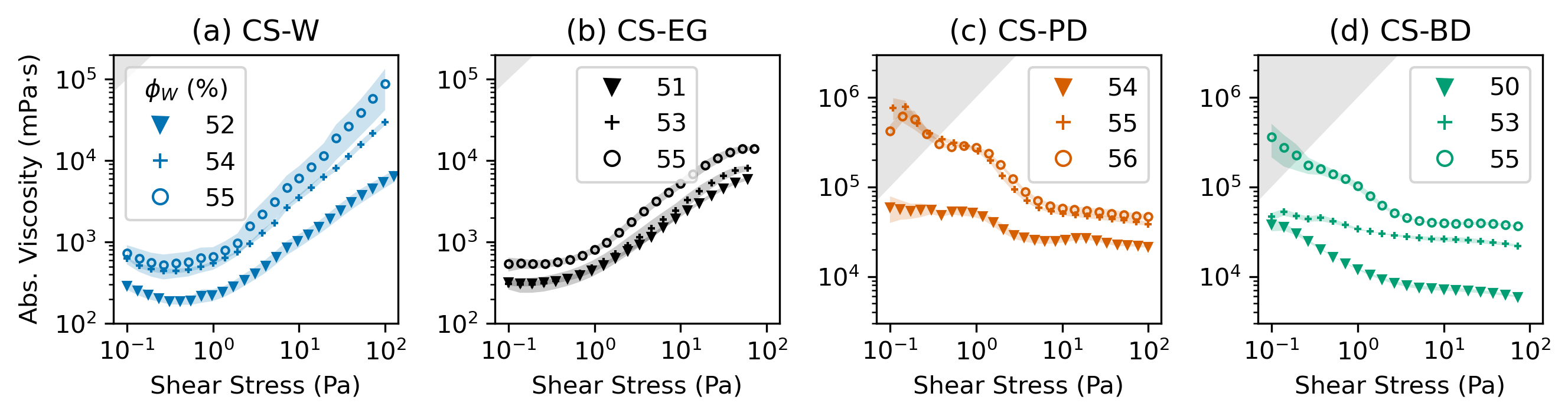} 
    \caption{
    Stress-controlled rheology for suspensions of CS in (a)~water, (b)~EG, (c)~PD, and (d)~BD, showing absolute viscosity $\eta$ as function of applied shear stress $\tau$ for different particle weight fractions $\phi_{W}$.
    Each point is the mean of three distinct trials; the colored shaded regions indicate the standard deviation calculated from a set of three trials.
    The gray regions indicate regimes that lie outside the standard operational limits of the rheometer.
    }
    \label{fig:rheology}
\end{figure*}

The stress-dependent viscosity of cornstarch (CS) particles in water is shown in Fig.~\ref{fig:rheology}a.
At low shear stress, the suspension exhibits Newtonian or mild shear-thinning behavior.
As shear stress increases past the onset stress for shear-thickening, $\tau_0$, the viscosity rises and the suspension begins to shear thicken. 
By increasing the particle weight fraction $\phi_{W}$, this strong thickening is exacerbated into DST, i.e., the shear thickening index $\beta = d\log\eta/d\log\tau \rightarrow 1$.
Measuring rheology at a higher stress levels $\tau>100$ Pa was limited by the onset of surface instability and ultimately sample ejection \cite{Maharjan_OReilly_Postiglione_Klimenko_Brown_2021,Garland_Gauthier_Martin_Morris_2013}.

Cornstarch suspensions in ethylene glycol (CS-EG) (Fig.~\ref{fig:rheology}b) also exhibit strong thickening, although with a smaller index $\beta$ compared to CS in water for a fixed $\phi_W$.
The EG suspensions did not reach DST even at $\phi_{\textrm{W}}=55\%$, where $\beta=0.8$. 
We associate this difference with a slight reduction in the particle-solvent H-bonding strength between EG and CS as compared to W and CS. 
However, the size difference between the two solvent molecules appears to be too small to noticeably affect the onset stress for thickening, which is $\tau_0\approx 1$~Pa in both cases, suggesting that the interparticle repulsive force $F_\mathrm{R}$ is comparable in both suspending mediums \cite{krishnamurthy2005shear,guy2015towards}.

In moving from EG to propanediol (PD) suspensions, the number of carbon atoms in the solvent backbone is increased from two to three, and this small change dramatically reduces the strength of particle-solvent H-bonding. 
The rheology of CS-PD suspensions in Fig.~\ref{fig:rheology}c shows a stark difference compared to the shorter solvents, exhibiting much higher viscosity at low $\tau$ and shear-thinning as $\tau$ increases. 
No thickening behavior is observed within the measured $\tau$ range.
Adding a fourth carbon to the solvent molecules by using butanediol (BD) is found to result in shear thinning rheology qualitatively similar to what we see with PD (Fig.~\ref{fig:rheology}d).
Both CS-PD and CS-BD show behavior similar to what is found for CS suspensions in non-aqueous solvents \cite{oyarte_galvez_dramatic_2017,Richards_Guy_Blanco_Hermes_Poy_Poon_2020}, where shear thinning can be associated with the formation of particle aggregates and a gel-like microstructure in the quiescent state, which gives rise to a yield stress \cite{Raghavan_Walls_Khan_2000}.

As the number of carbon atoms increases, the decrease in solvent polarity (reflected by the changes in dielectric constant $\varepsilon_\mathrm{r}$, see Table~\ref{tab:solvents}), is expected to result in a weakening of the solvation layer surrounding the particles.
This decrease then leads to less favorable particle wetting conditions, and we believe that this is, at least in part, responsible for the weakening of the shear thickening we observe in moving from water to EG as the solvent. In principle, further reduction of $\varepsilon_\mathrm{r}$ and thus wettability will eventually lead to particle aggregation by reducing interparticle steric repulsion.
However, the striking difference we see in Fig.~\ref{fig:rheology} between EG and PD is not simply a consequence of a reduction in $\varepsilon_\mathrm{r}$: if the dielectric constant of EG is similarly reduced by changing temperature, no change in the character of the shear thickening behavior is found (see Supporting Information and Figure~S2).
To further investigate these changes in solvation, in the following section we employ density functional theory simulations to understand the different molecular conformations that each solvent molecule can take.

\begin{figure}[!b]
    \centering
    \includegraphics[width=0.5\linewidth]{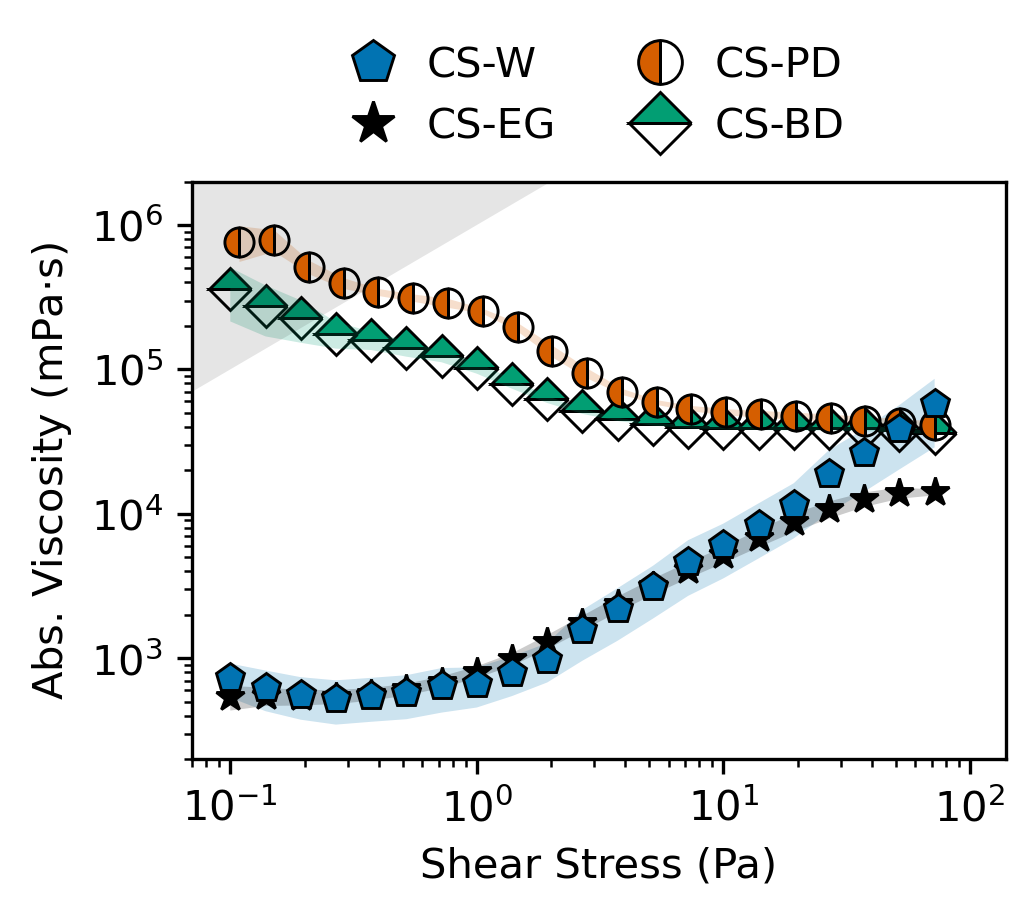}
    \caption{
    Viscosity as a function of shear stress for suspensions of CS in water, EG, PD, and BD at a weight fraction $\phi_{W}=55$\%.
    The gray region indicates the shear rates that lie outside the standard operational limits of the rheometer.
    }
    \label{fig:55wtpct}
\end{figure}

Previous work has shown that, generally, shear thinning behavior associated with an apparent yield stress tends to mask any underlying shear thickening, at least at low stress \cite{brown_shear_2014,Singh_Pednekar_Chun_Denn_Morris_2019}. 
However, when larger shear stress starts to activate frictional particle-particle contacts, we would expect the rheology to become largely independent of effects related to the solvent.
In other words, we would expect the behavior of all four suspension types to be dominated by friction between cornstarch particles for larger applied shear stresses.
At the upper end of the stress range accessible by our measurements of the shear rheology, we indeed find that the flow curves for all four suspensions with $\phi_{W}=55$\% approach each other (Fig.~\ref{fig:55wtpct}). 
Still, to probe whether there is convergence in behavior, we need to extend the stress range to much higher levels than possible with rheometry, and to that end we employ pull tests as described further below.

\subsection{Density functional theory for diol solvents}
\label{DST-sec}

\begin{figure}[ht]
    \centering
    \includegraphics[width=0.5\linewidth]{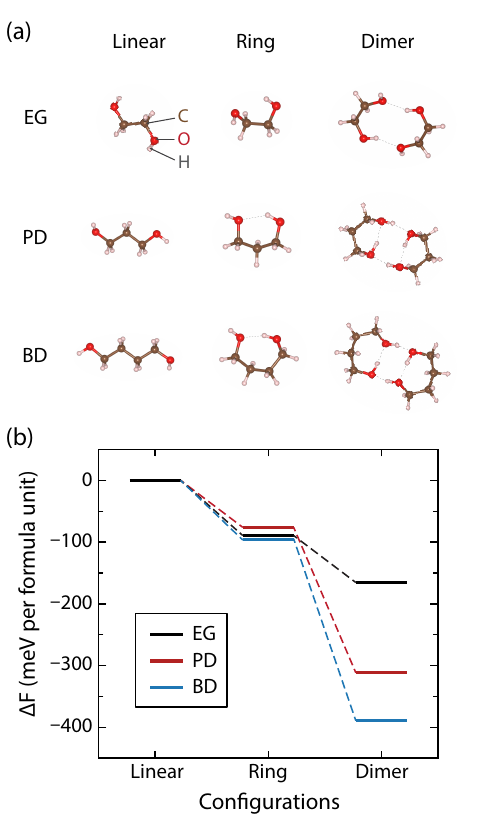}
    \caption{(a) Molecular structures of EG, PD, and BD in three configurations: linear, ring, and dimer. Carbon, oxygen, and hydrogen atoms are shown as brown, red, and pink spheres, respectively. (b) Relative vibrational free energies ($\Delta F$) of the three configurations at 300 K, computed by using Eq.~\ref{eq:Fvib} with B3LYP electronic energies.}
    \label{fig:DFT}
\end{figure}

To explain the dramatic change in the rheological behavior between EG and PD, the effect of carbon-chain lengths in diol molecules (EG, PD, and BD) is investigated by performing density functional theory calculations.
Since the surveyed diols share the same number of hydroxyl groups, we hypothesize that longer carbon chains suppress the formation of strong solvation layers around the CS particles.
In Fig.~\ref{fig:DFT}a, we show the optimized molecular geometries of the diols in three different configurations: (i) linear, (ii) ring, and (iii) dimer.
The ring and dimer configurations are stabilized by intra- and inter-molecular hydrogen bonding, respectively. Although the ring configuration of the EG molecule does not form a distinct intramolecular hydrogen bond, we classify its \textit{cis-}arrangement as a ring configuration based on geometric similarity.

In Fig.~\ref{fig:DFT}(b), we compare the thermodynamic stability of diol configurations based on vibrational free energies obtained from DFT.
For all diols, the ring and dimer configurations are found to be more stable than the linear arrangement.
While the energy differences between the linear and ring configurations are relatively small for diols (around 20\,meV/f.u.),
a notable shift in energy landscape occurs for dimer configurations.
Specifically, the energy merit of forming a dimer is significantly smaller for EG molecule than for PD and BD. The energy decreases from the ring to the dimer configuration by 75.7 meV/f.u. for EG, whereas the corresponding energy differences are 236 and 293 meV/f.u. for PD and BD molecules, respectively.
This tendency can be attributed to the fact that longer chain length reduces the energy penalty when stretching to form the dimer conformation.
We note that this energy landscape is robust for different functionals (see Supporting Information and Figure~S3).

As a result, compared to EG, a larger fraction of PD and BD solvent molecules likely exists in the dimer form, reducing the number of free H-bond acceptors and donors in the bulk suspension.
This reduction suppresses the formation of solvation layers around CS particles thereby weakening interparticle repulsion. 
Consequently, CS particles suspended in PD and BD solvents exhibit yielding and shear-thinning behavior.
This connection between atomic-scale conformation and macroscopic rheology shows that the non-Newtonian behavior depends not only on particle-solvent but also on solvent-solvent interactions.

\subsection{Probing the shear-jammed regime}
\label{sec-pull-test}

\begin{figure*}[htpb]
    \centering
    \includegraphics[width=0.9\linewidth]{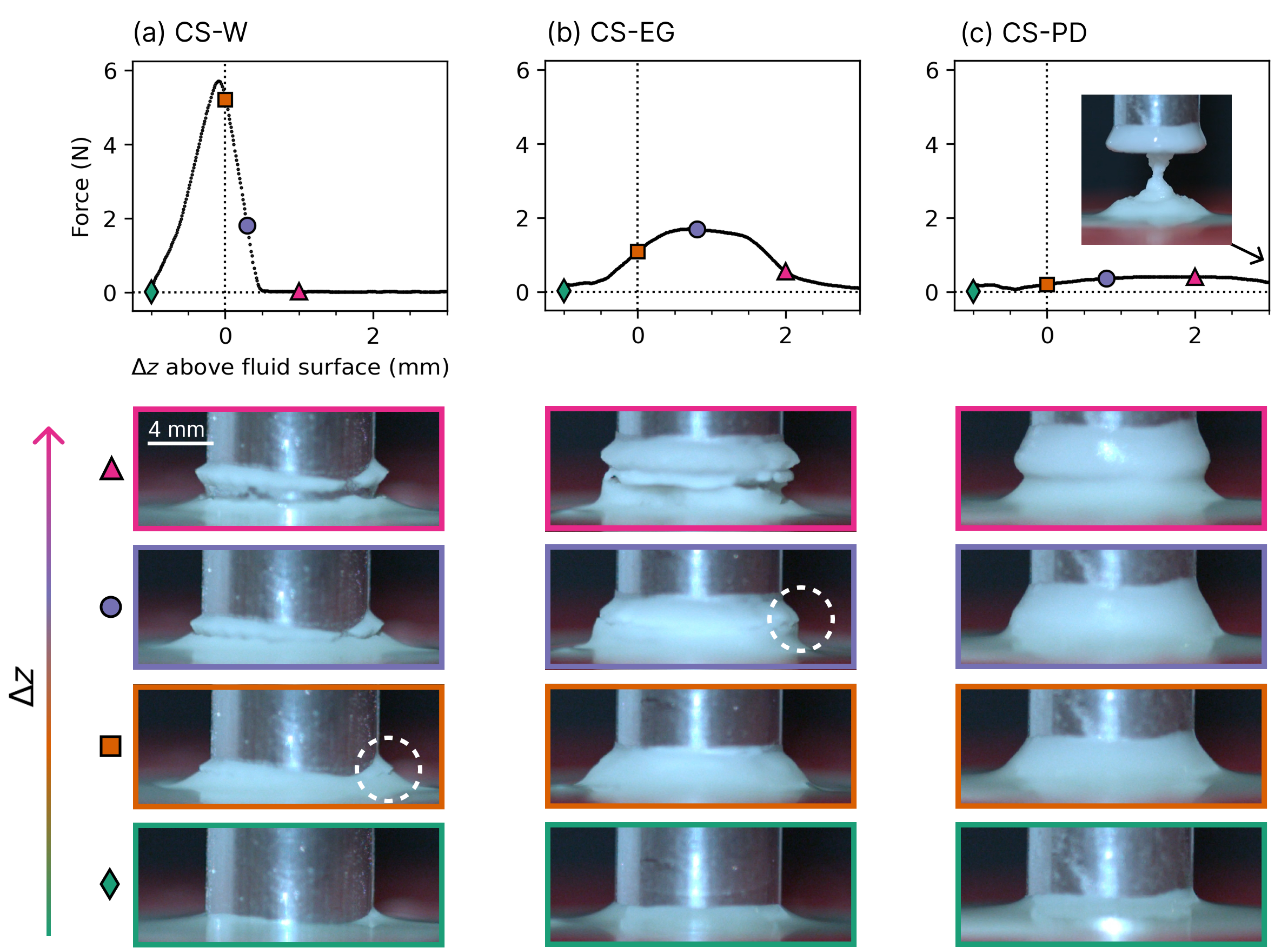}
    \caption{Representative force curves and video stills from tensile tests at $v=10$~mm/s for (a) CS-W, (b) CS-EG, and (c) CS-PD suspensions of weight fraction $\phi_{\textrm{W}}=55$\%.
    White dashed circles indicate first visual sign of suspension fracture. Image inset in (c):  pronounced necking for CS-PD at larger extension.
    }
    \label{fig:Fig6}
\end{figure*}

\begin{figure*}[htpb]
    \centering
    \includegraphics[width=0.8\linewidth]{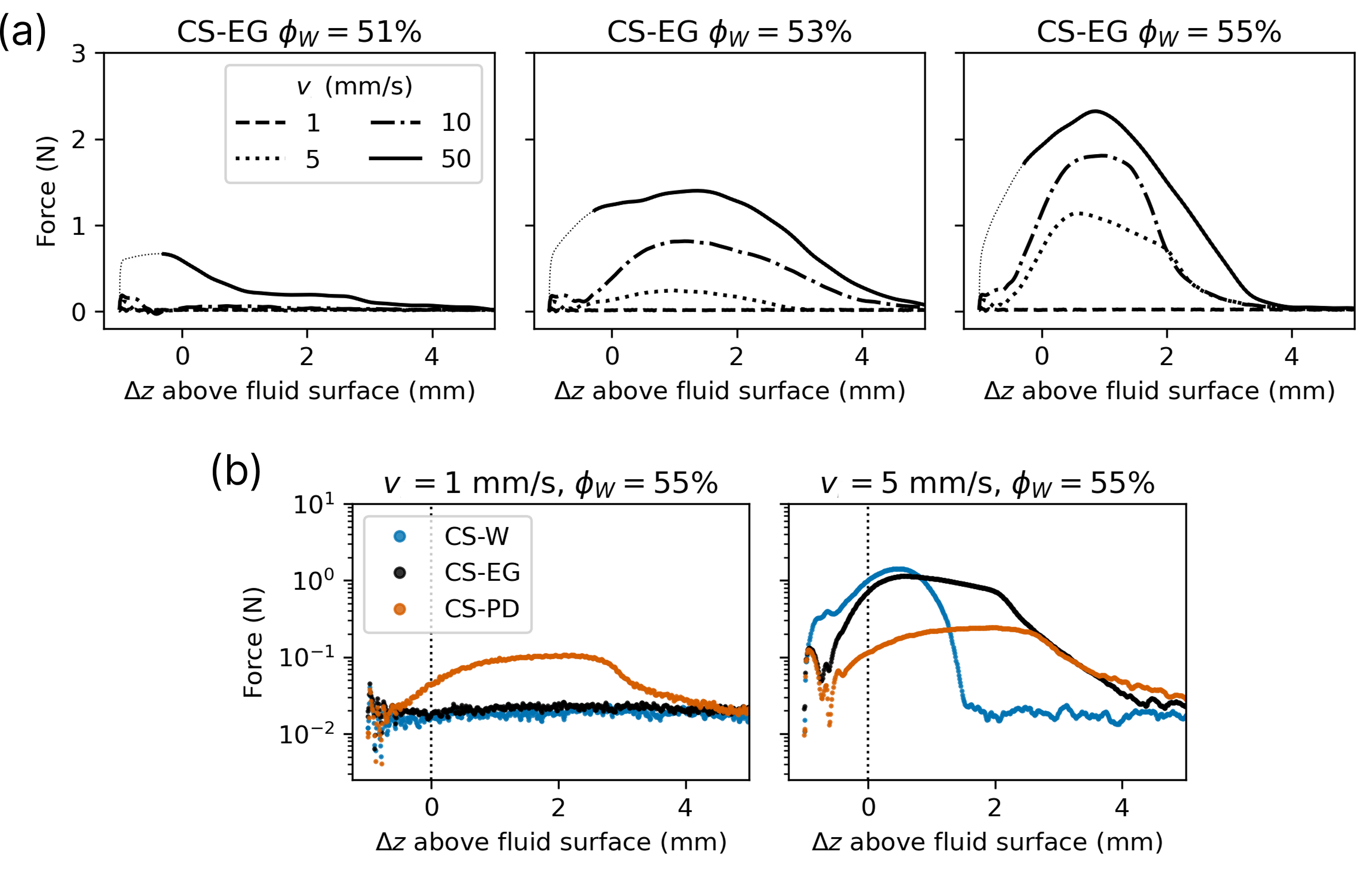}
    \caption{
    (a) Force versus displacement curves of CS-EG suspensions for varied weight fractions and pulling speeds $v$.
    At least three pulls were performed for each $v$, but only one is plotted for clarity.
    For 50~mm/s the initial instrument acceleration was significant (see SI), affecting the measured force in the regions indicated by the faint dotted lines, and is not considered in our analysis.
    (b) Force traces for CS-W, CS-EG, and CS-PD $\phi_W=55$\% for $v=$1~mm/s (left) and 5~mm/s (right).
    }
    \label{fig:Fig5}
\end{figure*}

We probe the shear jamming regime with pull tests, applying uniaxial tensile strain to the suspension by moving an initially submerged rod upward at fixed speed $v$ along the vertical, $z$-direction while recording the force exerted on the rod. 
Figure~\ref{fig:Fig6} shows representative results for CS-W, CS-EG and CS-PD, obtained for $\phi_{\textrm{W}}=$ 55\% and speed $v=$ 10~mm/s. 
Similar results were obtained with $v=$ 50~mm/s (see Supporting Information Figure~S4).
Although the cross-section of material being pulled up by the rod changes with strain, we can estimate the applied stress during the initial stage around $\Delta z = 0$, by dividing the force by the area of the flat rod face. 
A measured force of 1~N thus corresponds to 12.7~kPa.

Beginning with CS-W (Fig.~\ref{fig:Fig6}a), the suspension sustains 1~mm of deformation (orange square) before two important visual changes occur: (1) the surface of the suspension transitions from shiny to matte, and (2) the meniscus formed by the suspension around the rod cracks, as indicated by the white dashed circle.
The observed surface reflectance change from shiny to matte is due to dilation and, together with the large measured force, is indicative of the transformation of the suspension material into a DST or SJ state \cite{Slutzky_2026,Brown_Jaeger_2012,Majumdar_Peters_Han_Jaeger_2017}.
As the rod continues to move upward, the crack propagates around the meniscus, and the suspension detaches from itself.
The formation of the initial crack corresponds to the steep reduction in measured force seen in Fig.~\ref{fig:Fig6}a.
Once the rod reaches $\Delta z=1$~mm, the surface of the suspension returns to a shiny reflectance.
This sequence of stages suggests that the suspension starts off in a liquid-like state, almost immediately is sheared into DST (matte appearance) and SJ (formation of brittle crack), and then relaxes back to a liquid-like state once the crack has separated the bulk of the fluid from the rod.

A similar 3-stage process is found for CS-EG (Fig.~\ref{fig:Fig6}b), although the force signal looks distinct from that of the CS-W case.
The suspension begins with a shiny surface, up to some critical deformation when the surface turns matte.
Soon after this change in appearance, a crack forms around $\Delta z\sim$1~mm, but the edge of the fracture is less jagged than for CS-W.
The fracture continues until the suspension completely separates from itself.
Note that the CS-EG suspension sustains more plastic deformation than CS-W before failing. 
This aligns with the shear rheology, where we also observe more rigidity from the CS-W suspension than from CS-EG.

In contrast to CS-W and CS-EG, the CS-PD suspension exhibits extensive ductile necking rather than brittle fracture, and as a result the force signal is both lower and extends to larger $\Delta z$ before complete failure occurs (Fig.~\ref{fig:Fig6}c). 
The images show the suspension pulling up with the rod and forming a cone of material before necking (see inset and also SI Movie 1, which extends to later times).
Despite the lack of clear fracture, the CS-PD sample undergoes the same shiny-matte-shiny sequence as a function of increasing strain, much like CS-EG and CS-W.
The matte appearance, along with the self-supporting cone formed under the rod, suggests that CS-PD also exhibits solid-like rigidity and shear jamming at these high stresses, albeit expressed very differently from CS-W and CS-EG.

The trends seen in Fig.~\ref{fig:Fig6} are preserved when the particle weight fraction $\phi_W$ or the pulling speed $v$ are varied. 
Figure~\ref{fig:Fig5}a shows representative force-displacement curves for CS-EG; see the SI for the full data set that also includes CS-W and CS-PD.
In general, smaller $\phi_W$ or $v$ weakens the response but does not significantly change the shape of the curves (at the lowest pulling speed, $v=1$~mm/s, we were limited by the force resolution of the instrument, which was around 20~mN (Fig.~\ref{fig:Fig5}b) and thus corresponds to roughly 250~Pa).
Conversely, increasing the pulling speed up to 50~mm/s is always seen to increase the measured stress response, implying that the suspensions exhibit strong thickening (and potentially jamming) at high levels of applied stress for all solvents tested, even if the low-stress rheology shows thinning behavior. 

Interestingly, at low pulling speeds and thus low applied stress, the force response of CS-W can be weaker than for the other solvents.
This is seen in Fig.~\ref{fig:Fig5}b by comparing CS-W, CS-EG, and CS-PD at 1~mm/s for $\phi_w=55$\%.
We associate this with the fact that water and EG form the strongest solvation layers around the CS particles, which provides repulsion that needs to be overcome for particle-particle H-bonding to occur.
Weaker solvation of particles, by using PD or BD as the suspending liquid, leads to weaker repulsion and enables onset of H-bonding at lower stresses.
Presumably, this bonding involves fewer and longer molecular bridges than with water, which leads to weaker particle-particle coupling.
We speculate that, at the same time, this weaker coupling between the particles makes the jammed solid less brittle and thus enables it to sustain larger tensile strain before failing. 
This idea is reinforced by the 5~mm/s data in Fig.~\ref{fig:Fig5}b: CS-W fails in a brittle fashion (rapid force drop at comparatively small extension), whereas CS-EG and CS-PD show more gradual decays in force corresponding to non-brittle failure.

\section*{Conclusions}
At low shear stress $\tau$, strongly solvated particles repel one another sterically via a solvent molecule barrier; at high $\tau$, these same solvent molecules can form bridges between particles.
As such, H-bonds can both hinder and enable frictional particle interactions.
A macroscopic analog may be found in common hook-and-loop fastener strips, which sterically repel each other at very low applied pressure, but form connections once a certain pressure threshold has been exceeded such that the hooks and loops get close enough to engage, i.e., to form reversible bridging bonds.

By varying the number density of solvent H-bonding sites, as well as the length of our solvent molecules, we effectively change the strength of interparticle repulsion at low applied shear as well as the strength of interparticle bridging at high applied shear.
This is directly observable in the shear rheology, where the suspensions using shorter solvents (W and EG) exhibit discontinuous and continuous shear thickening, respectively.
The longer solvent molecules (PD, BD) exhibit entirely contrasting shear rheology, where they begin as yield-stress fluids that shear thin with increasing $\tau$.
Density functional theory simulations reveal that increasing the length of the solvent molecule unlocks additional conformational complexity, where EG, PD, and BD molecules can form rings and dimers.
This self-interaction reduces the solvent's H-bonding availability for interacting with the solid particles, resulting in weaker solvation at low $\tau$ and longer, more flexible H-bond bridges between particles at high $\tau$.

We find that these solvent effects carry over into in the shear jamming regime.
Suspensions made with longer solvent molecules (PD) are largely softer under extension than their shorter counterparts (W, EG) which deform in a brittle manner.
Such a mechanically different behavior is observed not only in the measured tensile forces, but also by visual observations of their eventual mechanical failure, indicating that solvents modify frictional particle interactions via H-bonded molecular bridges even at very large applied stresses.

\section*{Author contributions}
\#H.K. and S.M.L. contributed equally. H.K. performed the project conceptualization, S.M.L. and H.K. performed experimental investigation, Y.S. conducted the DFT calculation, S.M.L., H.K., and H.M.J. developed the methodology, and all authors contributed to the data analysis and writing of the paper.

\section*{Conflicts of interest}
There are no conflicts to declare.

\section*{Data Availability}
All data supporting this study are provided as part of the article and its Supporting Information.

\section*{Acknowledgements}
We acknowledge support from the University of Chicago Materials Research Science and Engineering Center, which is supported by the NSF under award DMR-2011854. 
Additional support was provided by the Army Research Laboratory under Cooperative Agreement Numbers W911NF-20-2-0044 and W911NF-25-2-0156.
This research was partly supported by Korea Institute for Advancement of Technology (KIAT) grant funded by the Korea Government (MOTIE) (RS-2025-02214408, HRD Program for Industrial; Innovation).
The computational work was supported by the Supercomputing Center/Korea Institute of Science and Technology Information with supercomputing resources, including technical support (KSC-2025-CRE-0419).

\section*{Supporting information}
The following files are available free of charge.
\begin{itemize}
  \item File1: Additional shear rheology results, pull-test results, and DFT results.
  \item Movie-S1: 10~mm/s pull-test of $\phi_W=55$\% suspensions in water, EG, and PD.

\end{itemize}

\printbibliography

\newpage
\Large{\textbf{Table of Contents Entry}}\\
\rule{0.05in}{1.75in}%
\begin{minipage}[b][1.75in]{3.25in}
  \sffamily
  \frenchspacing

	\includegraphics[width = 3.25in]{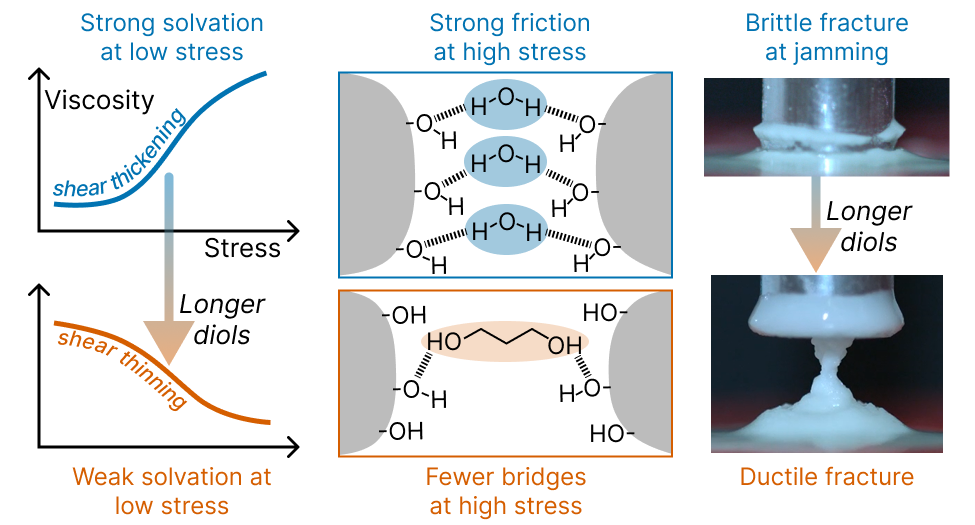}
\end{minipage}%
\rule{0.05in}{1.75in}

\end{document}